\documentstyle[12pt]{article}
\begin{document}

\begin{titlepage}
\pagestyle{empty}
\baselineskip=21pt
\rightline{hep-th/yymmddd}
\rightline{May 1997}
\vskip .2in
\begin{center}
{\large{\bf On Graceful Exit in String Cosmology\\ with Pre-Big Bang Phase }}
\end{center}
\vskip .1in
\begin{center}
Aram A. Saharian

{\it Department of Physics, Yerevan State University}

{\it 1 Alex Manoogian St., 375049 Yerevan, Armenia}

\end{center}
\vskip .5in

\centerline{ {\bf Abstract} }
\baselineskip=18pt

We analyze the problem of graceful exit from superinflationary pre-big bang
phase of string cosmology within the context of lowest-order string
effective action. The previous no go theorems are generalized for the case
when higher genus terms of general form and additional matter fields are
included. It is shown that the choice of the E-frame essentially simplifies
the consideration. For the example of pure gravi-dilaton case the comparison
of E-frame and string frame approaches, based on phase space analysis, is
carried out.

\bigskip
PACS Number(s): 98.80.Bp, 98.80.Cq, 98.80.Hw
\end{titlepage}
\baselineskip=18pt

\section{Introduction\ }

Inflationary cosmology (for reviews see \cite{infl} ) was proposed as a
possible solution to a number of outstanding problems of standard
cosmological model such as the horizon, flatness, space-time isotropy and
homogeneity, the structure formation, magnetic monopoles over-abundance
problems and so on. The inflation requires a phase of accelerated expansion
in the early universe and has been the subject of much investigation during
the past decade. However the final form of the model is not yet fixed. The
first proposed inflationary scenario, called old inflation \cite{oldinfl},
based on a first-order phase transition, could not provide a satisfactory
explanation of how to get out from the inflationary phase without disturbing
the good properties of the standard cosmological model \cite{Guth}. The new
inflationary scenario \cite{newinfl} was proposed to solve this graceful
exit problem with second order phase transition. The field slowly rolls down
the finite temperature effective potential at first, with exponential
expansion occurring. Inflation terminates when the field leaves the slow
rolling regime, quickly evolves to the true minimum, and reheats via
oscillations about the bottom of the potential. A different solution of
graceful exit problem, known as chaotic inflation was proposed in \cite
{chaotic}. This scenario showed that inflation need not occur only in very
special field theories. As in the case of new inflationary scenario, here
the density fluctuations force the couplings to be excessively small and
this models do suffer from a fine-tuning problems.

Recently, great interest has been devoted to the study of extensions of the
inflationary scenario based on scalar-tensor theories of gravity. La and
Steinhardt \cite{extinfl} proposed a model, known as extended inflation,
based again on a first-order phase transition, where graceful exit problem
was solved by using Jordan-Brans-Dicke (JBD) theory. The crucial feature of
these models is that their inflationary solutions are power-law rather than
exponential \cite{powerlaw} (power-law inflationary expansion also can arise
in theories of minimal gravity and an exponential scalar potential \cite
{exppot}). Unfortunately homogeneity afterwards is achieved only for JBD
parameters which violate observations \cite{extinfl1}. A possible way to
avoid this conflict is the so-called hyperextended inflation, based on more
general scalar-tensor theories \cite{hypext}. The other possibility for
successful extended inflation might be multidimensional theories \cite
{multinf}. The relationship between various theories of inflation is
examined in \cite{kalara}.

Inflationary models in general require small parameters in particle theory
Lagrangian, to provide the flat potential needed for sufficient inflation
and for correct magnitude of density fluctuations. The models of inflation
with no unmotivated small parameters can be constructed by allowing for more
than one field to be relevant to inflation as in ''hybrid inflation'' \cite
{hybridinfl}, soft inflation \cite{softinf}, and supernatural inflation \cite
{natinf} models.

Recently, a great deal of attention has been devoted to possible
implementations of the inflationary scenario in supergravity/superstring
models (see, for example, \cite{supinf, sah1} and references therein). The
moduli fields, which parametrize perturbative flat directions of the
potential in supersymmetric theories, are natural candidates to act as
inflatons. An interesting alternative to the standard inflationary universe,
motivated by the scale factor duality of the string effective action \cite
{sdual}, has been developed in \cite{gasp1, gasp2, gasp3}. In this scenario
(generically referred to as pre-big bang cosmology) the evolution starts
when the string dilaton is deep in the weak coupling region and Hubble
parameter is small. The evolution in this epoch is an accelerated expansion
dominated by the dilaton kinetic energy and determined by the vacuum
solution of the string gravi-dilaton equations of motion (kinetic inflation
was also discussed in \cite{levin}). It is assumed that after a period of
time, of length determined by the initial conditions, a branch change, or
phase transition, from the accelerated expansion phase occurs (pre-big bang
phase) into a phase which will eventually become a phase of decelerated
expansion (post-big bang phase). However, confirming a previous conjecture 
\cite{brust} it has been shown \cite{cal1} that such a branch change can not
occur for a realistic dilaton potential if one is limited to the lowest
order expansion of the string theory. Subsequently this result has been
extended to the more general cases of gravi-dilaton-axion system with
axion-dilaton potential \cite{cal2} and for models with spatial curvature 
\cite{east}. On the other hand, a quantum cosmological approach based on the
tunnelling boundary condition results in a non-zero transition probability
from a pre-big bang to a post-big bang classical solution \cite{quantcos}.
Such a transition may be interpreted as a spatial reflection of the
wavefunction in minisuperspace. In \cite{higher} it has been shown that
quantum corrections arising in the strong coupling regime can regularize the
curvature singularity of the tree-level pre-big bang models.

In the present paper we will discuss the graceful exit from pre-big bang
phase to lowest order in the $\alpha^{^{\prime }}$ expansion of the string
effective action for the case when higher genus terms of general form and
additional matter fields are included. In the next section, the structure of
the lowest order effective action is considered in both string and Einstein
frames, and equations of string cosmology are derived. In Section 3 we
briefly outline the pre-big bang scenario and examine the possibility of
having a branch change from the accelerated expansion phase into the
decelerated post-big bang one. The Section 4 is devoted to phase space
analysis and the relation of Einstein and string frame approaches is
discussed. Finally in Section 5 we summarize our main results.

\section{String Effective Action and Cosmological Equations}

Perturbative string theory contains two parameters: the string tension $%
\alpha ^{^{\prime }}$ of inverse mass-squared dimension and dimensionless
string coupling constant $g_s$ . The first one sets the length scale of the
theory and controls stringy effects in the sense that when $\alpha
^{^{\prime }}\rightarrow 0$ the theory becomes equivalent to a field theory.
The second parameter $g_s$ controls quantum effects and plays the role of
loop expansion parameter. At lowest order in $\alpha ^{^{\prime }}$ the
string effective action reads \cite{staction}

\begin{eqnarray}
S &=&-\frac 1{16\pi G_D}\int d^Dx\sqrt{\left| G\right| }[F_R(\varphi
)R+4F_\varphi (\varphi )\partial _M\varphi \partial ^M\varphi -\frac
1{12}F_H(\varphi )H^2+  \nonumber \\
&&\qquad +V(\varphi )-16\pi G_DL_m(\varphi ,G_{MN},\psi )]  \label{effact}
\end{eqnarray}
where $R$ denotes the curvature scalar of the $D-$ dimensional metric $%
G_{MN} $, $\varphi $ is the dilaton field, $H^2=H_{MNP}H^{MNP}$, $%
H_{MNP}=3\partial _{[P}B_{MN]}$ is the Kalb-Ramond field strength. The
string coupling constant is related to the expectation value of the dilaton
as $g_{s=}\langle e^{2\varphi }\rangle $ . The last term in (\ref{effact})
denotes the Lagrangian density of other fields, collectively denoted by $%
\psi $ , which is a function of the metric and dilaton field. As an example
we shall consider the case of gauge field. For the heterotic string

\begin{equation}
L_m=\sum \frac{F_g(\varphi )}{4g_i^2}F_{iMN}^aF_i^{aMN}  \label{lgauge}
\end{equation}
where $F_{iMN}^a$ is the gauge field strength, the index $i$ labels the
various simple components of the gauge group, while $a$ spans the
corresponding adjoint representations. The tree-level couplings 
$g_{i}$ are given by $g_i=1/\sqrt{k_i}$where $k_{i}$ are the
integer levels of the appropriate affine algebras responsible for the gauge
group.

In (\ref{effact}) we have introduced a potential $V(\varphi )$ for the
dilaton, which is expected to be non perturbative, related to supersymmetry
breaking. At small couplings ($\varphi \rightarrow -\infty $) it has to go
to zero as a double exponential $\exp (-\sigma \exp (-2\varphi ))$, with
some model dependent positive constant $\sigma $.

The functions $F_k(\varphi )$ , receive string perturbative as well as
non-perturbative corrections. They have the perturbative expansion 
\begin{equation}
F_k(\varphi )=e^{-2\varphi }\left[ 1+\sum Z_k^{(i)}e^{2i\varphi }\right] ,
\label{expan}
\end{equation}
where dimensionless coefficients $Z_k^{(i)}$ represent the $i$ - loop
correction. The action (\ref{effact}) is written in the so-called string
frame metric to which test strings are directly coupled. For many purposes
it is more convenient to work with Einstein frame action. Upon performing a
conformal transformation given by ( further we will assume that the function 
$F_R(\varphi )$ is positive for all values of dilaton) 
\begin{equation}
G_{MN}=\Omega ^2(\varphi ){\bar G}_{MN},\qquad \Omega ^2(\varphi
)=F_R^{-2/(D-2)}  \label{trans}
\end{equation}
we can obtain the effective action in Einstein frame (hereafter referred to
as E-frame) where the pure gravitational action takes the standard
Einstein-Hilbert form: 
\begin{eqnarray}
S &=&\int d^Dx\sqrt{\left| G\right| }\{-\frac 1{16\pi G_D}[\bar R+4F\bar
G^{MN}\partial _M\varphi \partial _N\varphi -\frac 1{12}\bar F_H(\varphi
)\bar H^2+  \label{eact} \\
&&+\bar V(\varphi )]+\bar L_m(\varphi ,\bar G_{MN},\psi )\}  \nonumber
\end{eqnarray}
Here the following notations are introduced ( a prime denotes
differentiation with respect to $\varphi $)

\begin{eqnarray}
F(\varphi ) &=&\frac{F_\varphi }{F_R}-\frac n{n-1}\left( \frac{F_R^{^{\prime
}}}{2F_R}\right) ^2,\quad \bar F_H=F_H\Omega ^{D-6},\quad n=D-1  \nonumber \\
\bar V(\varphi ) &=&\Omega ^DV(\varphi ),\quad \bar L_m=\Omega ^DL_m(\varphi
,\Omega ^2\bar G_{MN},\psi )  \label{notat}
\end{eqnarray}
If the function $F_R$ reverses the sign at some point, then in regions with
negative valued $F_R$ the transformation similar to (\ref{trans}) can be
performed with absolute value of function $F_R(\varphi )$. In this case the
Ricci scalar enters in (\ref{eact}) with positive sign, which corresponds to
negative sign gravitational constant. Note that at points where $F_R(\varphi
)=0$ the conformal transformation to the E-frame is singular.

By defining the conformal field $\phi $ as

\begin{equation}
\phi =\phi (\varphi ),\quad \left( \frac{d\phi }{d\varphi }\right)
^2=-4\beta F,\qquad \beta =-sgnF  \label{newphi}
\end{equation}
the pure gravi-dilaton Lagrangian density takes the form

\begin{equation}
L_{G\varphi }=\frac 1{16\pi G_D}\left[ -R+\beta (\partial \phi )^2-\bar
V\right]  \label{elag}
\end{equation}
In this paper we shall assume the dilaton field is non-tachionic, and
therefore $\beta >0$. At the tree-level for the functions $F_k(=e^{-2\varphi
})$ we have

\begin{equation}
F=-\frac 1{n-1},\qquad \phi =\sqrt{\frac 4{n-1}}\ \varphi  \label{treephi}
\end{equation}
and this is indeed the case.

Let us consider $D$-dimensional homogeneous and isotropic metric background
of Friedman - Robertson - Walker type, with time - dependent dilaton. The
string frame metric is given in terms of the lapse function, $N(t)$, and
scale factor $a(t)$:

\begin{equation}
ds^2=N^2(t)dt^2-a^2(t)dl^2,  \label{interval}
\end{equation}
where $dl^2$ is the metric on a $n$-space of constant curvature $k(=0,\pm 1)$%
. Introducing this ansatze into the action (\ref{effact}) yields, after
integrating over space and dividing by the space volume:

\begin{eqnarray}
S_{eff} &=&\int dt\ Na^{D-1}\{-\frac 1{16\pi G_D}[n(n-1)F_R\left( \frac{h^2}{%
N^2}+\frac{f^{^{\prime }}h\dot \varphi }{N^2}-\frac k{a^2}\right) + 
\nonumber \\
&&\qquad +4F\frac{\dot \varphi ^2}{N^2}+V(\varphi )]+L\}  \label{cosact}
\end{eqnarray}
where the dots denote time derivative, $h=\frac{\dot a}a$ is the Hubble
parameter and we have introduced the notation

\begin{equation}
L=\frac{F_HH^2}{192\pi G_D}+L_m,\qquad f(\varphi )=\frac 2{n-1}\ln (F_R)
\label{lag}
\end{equation}
The equations of motion in the $N=1$ gauge are the following:

\begin{eqnarray}
\ddot \varphi &=&-y\dot \varphi -\frac{F^{^{\prime }}}{2F}\dot \varphi ^2+%
\frac{e^f}{8F}\bar V^{^{\prime }}(\varphi )-\frac{2\pi G_D}{F_RF}(\frac{%
f^{^{\prime }}}2T+\alpha )  \nonumber \\
\dot h &=&-yh-\frac{f^{^{\prime }}b^{^{\prime }}}{2b}\dot \varphi ^2-k\frac{%
n-1}{a^2}+  \nonumber \\
&&\qquad +\frac{e^f}2\left[ \frac{-f^{^{\prime }}}{8F}\bar V^{^{\prime
}}(\varphi )+\frac{2V}{n-1}\right] +\frac{8\pi G_D}{F_R}\varepsilon b_1
\label{coseq}
\end{eqnarray}
where

\begin{eqnarray}
y &=&nh+\frac 12(n-1)f^{^{\prime }}\dot \varphi ,\qquad \alpha =\frac 1{%
\sqrt{\left| G\right| }}\frac{\delta \sqrt{\left| G\right| }L}{\delta
\varphi }  \label{cosnot} \\
T_N^M &=&diag(\varepsilon ,...,-p,...),\quad T_{MN}=\frac 2{\sqrt{\left|
G\right| }}\frac{\delta \sqrt{\left| G\right| }L}{\delta G^{MN}},  \nonumber
\\
b_1 &=&\frac 1n+\frac{1-np/\varepsilon }{n-1}\left( \frac 1n-\frac{b^2}%
4\right) +\frac{\alpha f^{\prime }}{8F\varepsilon },\quad b^2=-\frac{n-1}{4F}%
f^{\prime 2}  \nonumber
\end{eqnarray}
Furthermore, extremizing action (\ref{cosact}) with respect to $N$ yields
the constraint equation:

\begin{equation}
\frac{16\pi G_D}{F_R}\varepsilon +e^f\bar V(\varphi )=n(n-1)\left[
(h+f^{^{\prime }}\dot \varphi /2)^2+k/a^2\right] +4F\dot \varphi ^2
\label{consteq}
\end{equation}
As a consequence of dilaton dependence of Lagrangian $L$ the corresponding
energy - momentum tensor is acted upon by a dilaton gradient force

\begin{equation}
\nabla _MT_N^M=-\alpha \partial _N\varphi  \label{contin}
\end{equation}
where $\nabla _M$ denotes the covariant derivative defined by the string
frame metric. Within the cosmological context this equation reads

\begin{equation}
\dot \varepsilon +nh(\varepsilon +p)+\alpha \dot \varphi =0  \label{doteps}
\end{equation}
In the case of vanishing potential and equation of state $p/\varepsilon
=const,\alpha /\varepsilon =const$ the general solution of anisotropic
multidimensional string cosmological models are considered in \cite{gasp1,
copel, sah1}.

We shall now consider cosmological equations in E-frame. The associated
metric is

\begin{equation}
ds^2=d\bar t^2-\bar a^2(\bar t)dl^2  \label{emet}
\end{equation}
According to (\ref{trans}) and (\ref{emet}) the times and scale factors in
string and E-frames are related by

\begin{equation}
dt=e^{-f/2}d\bar t,\qquad a=e^{-f/2}\bar a  \label{frames}
\end{equation}
The corresponding set of cosmological equations formally can be obtained
from (\ref{coseq}), (\ref{consteq}) by substituting $f=b=0$ and replacing $%
(a,t)\rightarrow (\bar a,\bar t)$:

\begin{eqnarray}
\frac{d^2\varphi }{d\bar t^2} &=&-(D-1)\bar h\frac{d\varphi }{d\bar t}-\frac{%
F^{^{\prime }}}{2F}\left( \frac{d\varphi }{d\bar t}\right) ^2+\frac{\bar
V^{^{\prime }}(\varphi )}{8F}-\frac{2\pi G_D}F\alpha  \label{eframeq} \\
\frac{dh}{dt} &=&-nh^2-k\frac{n-1}{a^2}+\frac V{n-1}+\frac{8\pi G_D}{n-1}%
(\varepsilon -p)  \nonumber \\
16\pi G_D\varepsilon &=&n(n-1)(h+k/a^2)+4F(d\varphi /dt)^2-\bar V(\varphi) 
\nonumber
\end{eqnarray}
If the Lagrangian density $L$ depends only on $G_{MN}$and not on its
derivatives, then E-frame energy density, pressure and function $\alpha $
are related to the string frame quantities through 
\begin{equation}
\bar \varepsilon =\Omega ^D\varepsilon ,\quad \bar p=\Omega ^Dp,\quad \bar
\alpha =\Omega ^D\left( \alpha -\frac{\Omega ^{^{\prime }}}\Omega T\right)
\label{esteps}
\end{equation}
Introducing the new scalar function $\phi (\varphi )$ according to (\ref
{newphi}) the first of equations (\ref{eframeq}) becomes

\begin{equation}
\frac{d^2\phi }{d\bar t^2}=-n\bar h\frac{d\phi }{d\bar t}-\frac 18\bar
V^{^{\prime }}(\varphi )-\frac{2\pi G_D}{\sqrt{\left| \bar G\right| }}\frac{%
\delta \sqrt{\left| \bar G\right| }L}{\delta \phi }  \label{phieq}
\end{equation}
Assuming that the dilaton dependence of Lagrangian density $L$ has the form

\begin{equation}
L=F_L(\varphi )\tilde L(G_{MN},\psi )  \label{lagphi}
\end{equation}
where $\tilde L$ is a function of conformal weight $\beta $: $\tilde
L(gG_{MN}, \psi)=g^\beta \tilde L(G_{MN},\psi ),$it is obtained that

\begin{equation}
\alpha =\frac{F_L^{^{\prime }}}{F_L}L,\qquad T=-(D+2\beta )L  \label{alf}
\end{equation}
Two important special cases are the Kalb - Ramond field with $\beta =-3$ and
gauge field with $\beta =-2$ (see (\ref{lgauge})). If the equation of state
has the simple form $p=\lambda \varepsilon $ and $L=\lambda _0\varepsilon $
with constants $\lambda $ and $\lambda _0,$ the integration of (\ref{doteps}%
) yields

\begin{equation}
\varepsilon a^{n(1+\lambda )}=const\left| F_L(\varphi )\right| ^{-\lambda _0}
\label{epsco}
\end{equation}
The analogous result can be also derived in E-frame.

\section{Pre-Big Bang Cosmology and\protect\\Graceful Exit Problem}

The constraint equation (\ref{consteq}) can be used to eliminate one of two
functions $h$ and $\dot \varphi $ from the cosmological equations. Here as
such a variable we choose $h$:

\begin{eqnarray}
h &=&-f^{^{\prime }}\varphi /2\mp  \nonumber \\
&&\ \mp [\left( -4F\varphi ^2+16\pi G_D\varepsilon /F_R+e^f\bar V(\varphi
)\right) /n(n-1)-k/a^2]^{1/2}  \label{hfunphi}
\end{eqnarray}
The solutions to cosmological equations belong to two branches according to
which sign is chosen. First let us consider gravi-dilaton case with a flat
space ($k=0$) . By using (\ref{hfunphi}) the set of cosmological equations
can be written in the form of second order autonomous dynamical system with
respect to variables $(\varphi ,x=\dot \varphi )$: 
\begin{eqnarray}
\dot \varphi &=&x  \nonumber \\
\dot x &=&\frac 12\left( f^{^{\prime }}-\frac{F^{^{\prime }}}F\right) x^2\pm
x\sqrt{\frac n{n-1}}\sqrt{-4Fx^2+e^f\bar V(\varphi )}+\frac{e^f}{8F}\bar
V^{^{\prime }}(\varphi )  \label{phixsys}
\end{eqnarray}
In the absence of potential the phase trajectories are defined by equation 
\begin{equation}
x=x_0\left( -e^f/F\right) ^{1/2}\exp \left( \pm \sqrt{\frac n{n-1}}\phi
sgnx_0\right)  \label{xfunphi}
\end{equation}
on the phase plane $(\varphi ,x)$, and by equation 
\begin{equation}
h=-\left( \frac{f^{^{\prime }}}2\pm 2sgnx_0\sqrt{\frac{-F}{n(n-1)}}\right) x
\label{hphi}
\end{equation}
on the phase plane $(\varphi ,h)$ , with $x_0$ being a constant of
integration. In E-frame the time dependence of this solution is given by 
\begin{equation}
\phi =const-sgnx_0\cdot \sqrt{\frac{n-1}n}\ln \left| \bar t\right| ,\quad
h=\frac 1{n\bar t}  \label{ephit}
\end{equation}
where $-\infty <\bar t<0$ for the upper sign and $0<\bar t<\infty $ for the
lower sign. The corresponding string frame solution can be derived from the
relations (\ref{xfunphi}),(\ref{hphi}) and (\ref{hfunphi}). At tree-level
for functions $F_k$ ($=e^{-2\varphi }$) we obtain (see, for example, \cite
{gasp3} ) 
\begin{equation}
2\varphi =const-(1\pm \sqrt{n}sgnx_0)\ln \left| t\right| ,\quad h=\mp \frac{%
sgnx_0}{\sqrt{n}}\frac 1t  \label{treephi1}
\end{equation}
where again $-\infty <t<0$ for the upper sign and $0<t<\infty $ for the
lower sign. For the case $t<0$ (upper sign in (\ref{hfunphi}) and (\ref
{phixsys})) this solution describes either accelerated inflationary
expansion and evolution from a flat and weakly coupled ($\varphi <<-1$)
universe or decelerated contraction and evolution towards weak coupling.
From (\ref{phixsys}) it can be seen that for this type of solutions
(following \cite{brust} we shall refer to it as (+) - branch) the minimum $%
\varphi =\varphi _0$ of dilaton potential $\bar V(\varphi )$is unstable
fixed point (focus or node depending on relative values of $\bar V(\varphi
_0)$ and $\bar V^{^{\prime }}(\varphi _0)$, see \cite{sahpot}), and
therefore for such a solution dilaton cannot be fixed by potential. As it
follows from (\ref{ephit}) in E-frame the trajectories of (+) - branch
describe decelerated contraction (see \cite{gasp1, gasp2} for the relation
between two frames and for a discussion of their physical equivalence).

For the case $t>0$ (lower sign in (\ref{hfunphi}),(\ref{phixsys})) the
solution (\ref{treephi1}) describes either decelerated expansion or
accelerated contraction depending on the sign of integration constant $x_0$
(see (\ref{xfunphi})). For this type of solution the minimum of potential $%
\bar V(\varphi )$ is a stable fixed point and they can be connected smoothly
to a standard Friedmann-Robertson-Walker decelerated expansion with constant
dilaton. In E-frame, trajectories corresponding to (--) branch solution,
describe decelerated expansion.

In pre-big bang scenario of string cosmology \cite{gasp1, gasp2, gasp3} the
pre- and post- big bang phases are realized by (+) and (--) branch
solutions, correspondingly. According to this scenario the expansion of the
universe starts at $t\rightarrow -\infty $ when dilaton is deep in weak
coupling region ($\varphi <<-1$) and the Hubble parameter is small. The
evolution in this epoch (pre-big bang phase) is determined by the vacuum
solution (\ref{treephi1}) of string gravi-dilaton equations with upper sign, 
$x_0>0$ and $t<0$. After this period of superinflation, driven by dilaton
kinetic energy, the universe enters into the stage where the effects of
non-trivial dilaton potential and higher curvature terms become important
and a branch change into a phase of decelerated expansion (post-big bang
phase) occurs. In the post-big bang universe the dilaton value must be
fixed, since variation of dilaton field leads to changes in masses and
coupling constants, which are strongly constrained by observations. This can
be realized by including dilaton potential and trapping the dilaton in a
potential minimum (for another mechanism of dilaton fixation by higher genus
terms see \cite{damour}). If this is the case, the post-big bang stage of
evolution is described by the (--) branch solution. In this context, one of
the main problems is related to the question of whether the two branches can
be smoothly connected to one another. This is the graceful exit problem of
the pre-big bang scenario.

To investigate the possible ways of graceful exit it is more convenient to
work in E-frame rather than in string frame. In E-frame the cosmological
evolution of the pre-big bang scenario looks as following. The universe
contracts from the initial state when dilaton is in weak coupling region,
Hubble parameter is small and scale factor is large. In this stage the
evolution is determined by the (\ref{ephit}) with $t<0$. After some period
of decelerated contraction the universe enters to the stage where non
trivial dilaton potential and higher curvature terms become important. In
this stage a branch change into a phase of decelerated expansion occurs. In
E-frame the graceful exit problem of pre-big bang string cosmology at the
classical level, corresponds to the possibility of a continuous
contraction/expansion transition. It can be easily seen that such a
transition cannot be simply catalyzed within the framework of the string
effective action (\ref{effact}). Indeed, since the function $\bar h$ has
different signs in pre- and post-big bang phases, then the continuous
transition between them suggests that $\bar h=0$ and $d\bar h/d\bar t>0$ at
some moment in branch changing region. But as it can be easily seen from (%
\ref{eframeq}) 
\begin{equation}
\frac{d\bar h}{d\bar t}=-\frac{8\pi G_D}{n-1}(\bar \varepsilon +\bar
p)-\frac 1{n-1}\left( \frac{d\phi }{d\bar t}\right) ^2+\frac k{\bar a^2}
\label{consteq3}
\end{equation}
As follows from this equation $d\bar h/d\bar t<0$ for 
\begin{equation}
\bar \varepsilon +\bar p\geq 0,k=-1,0  \label{cond1}
\end{equation}
and we conclude that branch change from pre-big bang phase to post-big bang
one can not occur within the framework of lowest-order string effective
action (\ref{effact}), if these conditions are fulfilled and 
\begin{equation}
F_R(\varphi )>0,F(\varphi )<0  \label{cond2}
\end{equation}
as it is assumed in above analysis. Moreover as it can be easily seen, the
only property of the potential $\bar V(\varphi )$, we have used, is its
independence on metric. Therefore the previous statement on branch change
impossibility is valid also for the case when the potential $\bar V$ depends
on other scalar fields (for example,on axion field, see below).

The impossibility of (+)/(--) transition can be also seen immediately in
string frame. From (\ref{hfunphi}) it follows that for a continuous
transition from one branch to the other it is necessary 
\[
\left[ \left( -4F\dot \varphi ^2+16\pi G_D\varepsilon /F_R+e^f\bar V\right)
/n(n-1)-k/a^2\right] ^{1/2}= 
\]
\begin{equation}
=\mp \left( h+f^{\prime }\dot \varphi /2\right) =\mp \bar he^{f/2}=0
\label{cond3}
\end{equation}
where the second equation is obtained from (\ref{frames}). At transition
point 
\begin{equation}
\frac d{dt}\left( h+f^{\prime }\dot \varphi /2\right) =e^{f/2}\frac{d\bar h}{%
d\bar t}  \label{cond4}
\end{equation}
and the function $h+f^{\prime }\varphi /2$ decreases for the conditions (\ref
{cond1}). Therefore, the branch change, if it takes place, must be from the
(--) to the (+) branch.

By combining the last equation of (\ref{eframeq}) with (\ref{consteq3}) one
finds 
\begin{equation}
n(n-1)\frac 1{\bar a}\frac{d^2\bar a}{d\bar t^2}=-8\pi G_D\left[ n\bar
p+(n-2)\bar \varepsilon \right] -(n-1)\left( \frac{d\phi }{d\bar t}\right)
^2+\bar V  \label{haw}
\end{equation}
As it follows from here, if the total energy-momentum tensor, including the
contribution of dilaton field, satisfies to strong energy condition, then
right hand side of (\ref{haw}) is negative and scale factor becomes zero at
some finite time moment. At contraction-expansion transition point (in
E-frame) we have to have $(1/a)(d^2a/dt^2)\geq 0$, which is not the case for
(\ref{haw}), when strong energetic condition is satisfied. In this context
the no-go theorem is the consequence of Hawking-Penrose theorem on
singularities \cite{hawking}.

Let us consider in more detail the case of Kalb-Ramond field as a source in
cosmological equations. For the case of $D=4$ the equation of motion of this
field 
\begin{equation}
\partial _M\left( \sqrt{\left| G\right| }F_H(\varphi )H^{MNP}\right) =0
\label{kalb}
\end{equation}
can be solved by Freund-Rubin ansatz 
\begin{equation}
H^{MNP}=\frac 1{\sqrt{\left| G\right| }}F_H^{-1}(\varphi )\varepsilon
^{MNPQ}\partial _QA  \label{freund}
\end{equation}
with pseudoscalar axion field $A$. The corresponding contribution to the
Lagrangian $L$ (see \ref{lag}) is 
\begin{equation}
\frac{F_H^{-1}}{32\pi G_D}\partial _MA\partial ^MA  \label{axion}
\end{equation}
which corresponds to the matter with equation of state $\varepsilon =p$,
where $\varepsilon >0$ if $F_H>0$. As it follows from here the above
formulated no-go theorem is valid for Kalb-Ramond field even for the case
when the potential depends on axion field.

The special cases of above formulated results for $D=4$ previously have been
considered

\begin{itemize}
\item  in \cite{brust},\cite{cal1}when 
\begin{equation}
F_R=F_\varphi =e^{-2\varphi },\quad H=0,\quad L_m=0  \label{prev1}
\end{equation}

\item  \cite{cal1} when 
\begin{equation}
F_R=F_\varphi =e^{-2\varphi },\quad H=0  \label{prev2}
\end{equation}
and stringy fluid sources (in (\ref{effact}) $L_m$ does not depend on
dilaton field) with equation of state $p=\gamma \varepsilon ,\gamma
=const>-1/3$ are present.

\item  \cite{cal2} when 
\begin{equation}
F_R=F_\varphi =B(\varphi )\quad {\rm Damour-Polyakov \ \ ansatz}  \label{prev3}
\end{equation}
\[
H=0,\quad L_m=0
\]

\item  \cite{cal2},\cite{east} when 
\begin{equation}
F_R=F_\varphi =e^{-2\varphi },\quad L_m=0  \label{prev4}
\end{equation}
and Kalb-Ramond field is present.
\end{itemize}

\section{Phase Space Analysis}

We have considered the graceful exit of pre-big bang cosmology in terms of
variables $(\varphi ,\dot \varphi )$. In previous investigations of this
problem (see, \cite{brust, cal1, cal2}) usually the set of variables $(\dot
\varphi ,h)$ is chosen. Here we shall consider the relation of this
approaches for the simple case of pure gravi-dilaton system with flat space (%
$k=0$). First let us consider the corresponding E-frame dynamical system on
phase plane $(\phi ,d\phi /d\bar t=X)$. It can be easily obtained from (\ref
{phixsys}): 
\begin{equation}
\frac{d\phi }{d\bar t}=X,\quad \frac{dX}{d\bar t}=\pm \sqrt{\frac n{n-1}}X%
\sqrt{X^2+\bar V(\phi )}-\frac 12\bar V^{\prime }(\phi )  \label{dynsys}
\end{equation}
with Hubble parameter 
\begin{equation}
\bar h=\mp \sqrt{\frac{X^2+\bar V(\phi )}{n(n-1)}}  \label{hubble}
\end{equation}
As has been previously noted the upper/lower sign corresponds to
pre/post-big bang phases. For (\ref{dynsys}) classically allowed region is
defined by 
\begin{equation}
X^2\geq -\bar V(\phi )  \label{alreg}
\end{equation}
If the dilaton potential takes the negative values on some interval of
dilaton field then it can be seen that the boundary of this region 
\begin{equation}
X=\pm \sqrt{-\bar V(\phi )}  \label{bound1}
\end{equation}
is a solution of system (\ref{dynsys}). As a simple example we shall
consider the case of quadratic potential (the phase space analyze for the
more realistic dilaton potentials, arising from gaugino condensation
mechanism, see \cite{sahpot}) 
\begin{equation}
\bar V(\phi )=M^2\left( \phi ^2-\phi _1^2\right)   \label{quadpot}
\end{equation}
taking negative values on interval $-\phi _1<\phi <\phi _1$. For $x_0\phi
\rightarrow +\infty (-\infty )$the phase trajectories of dynamical system (%
\ref{dynsys}) with upper (lower) sign are described by equation (see (\ref
{xfunphi})) 
\begin{equation}
X=x_0\exp \left( \pm \sqrt{\frac n{n-1}}\phi sgnx_0\right) ,\quad x_0\phi
\rightarrow \infty   \label{traj}
\end{equation}
In addition, there are special trajectories with 
\begin{equation}
X=\pm M\sqrt{\frac{n-1}n}sgn\phi ,\quad \phi \rightarrow \infty 
\label{sptraj}
\end{equation}
For these trajectories $X^2<<\bar V(\phi )$ and they are potentially
dominated. Note that the straight lines (\ref{sptraj}) are exact solutions
of (\ref{dynsys}) if 
\begin{equation}
\phi _1=\phi _{10}=\sqrt{\frac{n-1}n}  \label{exsol}
\end{equation}
The corresponding time dependences of E-frame scale factor and scalar field
are given by 
\begin{equation}
\bar R=R_0\exp (-M^2\bar t^2/2n),\quad \left| \phi \right| =\pm M\sqrt{1-1/n}%
\bar t,\quad 0\leq \pm \bar t<\infty   \label{scalephi}
\end{equation}
This solution is non-singular everywhere.

The qualitative structure of phase portraits of dynamical system (\ref
{dynsys}) with dilaton potential (\ref{quadpot}) changes dependening on
which of the following intervals the constant $\phi _1$ lies: 
\begin{equation}
(a)\quad 0<\phi _1<\phi _{10},\quad (b)\quad \phi _1=\phi _{10},\quad
(c)\quad \phi _1>\phi _{10,}  \label{cases}
\end{equation}
where $\phi _{10}$ are defined by (\ref{exsol}). The phase portrait
corresponding to the first of this cases is presented in Fig.1. The
classically forbidden region is shaded (for the potential (\ref{quadpot})
the boundary of this region is ellipse). Solid/dashed lines correspond to
the E-frame expansion/contraction models (lower/upper sign in (\ref{dynsys}%
)). The special solutions (\ref{sptraj}) are presented by nearly horizontal
trajectories (see, for example, $A_1B_1$). The trajectories corresponding to
the lower sign in (\ref{dynsys}) (solid lines) are attracted to the boundary
(\ref{bound1}), touche it and then the branch change into the solutions with
upper sign (dashed lines) takes place. The last ones are repelled from the
boundary. As we see the branch change always occurs from expansion to
contraction phases. If the touching point of trajectory is the above (below)
one of special trajectories with $X>0$ ($X<0$) then for corresponding
solutions dilaton is monotonic increasing (decreasing) function of time. For
trajectories touching the boundary (\ref{bound1}) between the special
trajectories (\ref{sptraj}) with $X>0$ and $X<0$ the time derivative $\dot
\phi $ reverses the sign at some finite moment.

For the case (b) the touching points of special trajectories coincide with
the top of boundary. These trajectories are horizontal straight lines and
correspond to solutions (\ref{scalephi}). They are the only solutions with
monotonic behavior of dilaton field. Finally, for the case (c) of (\ref
{cases}) the touching points of special trajectories, coming from region
with $\varphi =-\infty $ ($\varphi =+\infty $), are on positive (negative)
half phase plane and there are no trajectories with monotonic dilaton field.

We shall now consider the phase trajectories of gravi-dilaton isotropic
models on phase space $(\dot \varphi ,h)$. Substituting from (\ref{consteq}) 
\begin{equation}
f_1\dot \varphi =-f^{\prime }h\pm \frac 4{\sqrt{n(n-1)}}\sqrt{-Fh^2+\frac
18f_1e^f\bar V},\quad f_1=\frac{8F}{n(n-1)F_R}  \label{dynsys2}
\end{equation}
into (\ref{eframeq}) we obtain the following set of equations in E-frame 
\begin{eqnarray}
\frac{d\phi }{d\bar t} &=&\pm \sqrt{(n-1)n\bar h^2-\bar V}  \label{dynsyse}
\\
\frac{d\bar h}{d\bar t} &=&-n\bar h^2+\frac{\bar V}{n-1}
\end{eqnarray}
We see that if potential is nonnegative everywhere then expansion and
contraction models are separated by classically forbidden region 
\begin{equation}
\bar h^2<\bar V/n(n-1)  \label{forreg}
\end{equation}
and there are no mixed expansion-contraction models. In particular, as a
necessary condition for successful branch change in pre-big bang cosmology
we obtain the existence of intervals with negative valued potential.

The phase portrait of dynamical system (\ref{dynsyse}) for the potential (%
\ref{quadpot}) with (\ref{cases}) (a) is shown in Fig.2, where the region (%
\ref{forreg}) is shaded. The solid/dashed lines correspond to the
upper/lower sign in (\ref{dynsyse}). The ellipse $AB_1B$ of Fig.1 is mapped
to the interval $AB$ with $-\phi _1<\phi <\phi _1$. The special solutions (%
\ref{sptraj}) are presented by trajectories passing through points $B_1$ and 
$B_2$. The trajectories with initial expansion will stop expanding when $h$
hits zero in the interval $AB$ and $h$ will then change the sign and the
universe will begin to contract. The trajectories intersecting the $\phi $
axis between $B_1$ and $B_2$ present the models with monotonic dilaton. The
other trajectories reach to the boundary of classically forbidden region (%
\ref{forreg}) (in horizontal direction) and then reflect from it. At that
point $d\phi /d\bar t$ reverses the sign and the trajectories described by
upper (lower) sign in (\ref{dynsyse}) have to turn into ones with lower
(upper) sign.

To compare the analysis carried out above with previous investigations of
graceful exit problem in pre-big bang string cosmology, we now turn to the
case of branch change in string frame. To be consistent with previous works
we should denote the trajectories having the plus/minus sign in front of
square root of (\ref{dynsys2}) as (+)/(--) branch trajectories. At tree
level for the functions in (\ref{expan}) we have 
\begin{equation}
f(\varphi )=-\frac{4\varphi }{n-1},\quad f_1=\frac 8{n(n-1)}
\label{treefunc1}
\end{equation}
and therefore from (\ref{dynsys2}) 
\begin{equation}
2\dot \varphi =nh\pm \left[ nh^2+\exp \left( -\frac{2\varphi }{n-1}\right)
\bar V(\varphi )\right] ^{1/2}  \label{phidot1}
\end{equation}
according to which the (+)/(--) branch trajectories are chosen in \cite
{brust}. From (\ref{dynsys2}) it follows that for a continuous (+)/(--)
transition it is necessary that 
\begin{equation}
h^2=\frac{f_1}{8F}e^f\bar V  \label{hub2}
\end{equation}
at a transition point. This equation defines the branch change curve. By
using the relation of Hubble parameters in string and E-frames 
\begin{equation}
he^{-f/2}=\bar h-\frac 12f^{\prime }(\phi )X  \label{hubrel}
\end{equation}
we obtain the equation of branch change curve in phase plane $(\phi ,\bar h)$%
: 
\begin{equation}
\bar h=\pm \frac{2sgnf^{\prime }}{n(n-1)}\sqrt{2F\bar V/f_1}\equiv \pm
sgnf^{\prime }\bar h^{(0)}  \label{hubeint}
\end{equation}
In Fig.2 this curve is shown by dash-dot line for the case $f_1>0$ when the
branch change occurs in region with negative valued potential (note that $%
F<0 $). At first look one would think that because the trajectories
intersect the curve (\ref{hubeint}) twice then the branch change will not
occur when all is said and done. However it is not difficult to see that
this is not the case. By taking into account that at transition point we
have to have $f_1\dot \varphi +f^{\prime }h=0$ from (\ref{hubeint}) one
obtains 
\begin{equation}
X=\pm \sqrt{\bar V\left( 4/nb^2-1\right) }\equiv \pm X^{(0)}  \label{xtrans1}
\end{equation}
at these points, where the upper (lower) sign corresponds to upper (lower)
sign in (\ref{hubeint}) and the relation 
\begin{equation}
f_1=(f^{\prime 2}/2)\left( 1-4/nb^2\right)  \label{relf1}
\end{equation}
is used. As we see, the (+)/(--) branch change occurs only when the
trajectories with $X>0$ intersect the (+) sign half of (\ref{hubeint}) or
when the trajectories with $X<0$ intersect the (--) sign half. Here we shall
consider the case when the function $f^{\prime }$ never changes the sign.
Since in weak coupling region $f^{\prime }<0$ (see (\ref{treefunc1})) this
suggestion means that this derivative is negative for all values of dilaton
field. Now we can see that in Fig.2 the trajectories in region $\bar h<-\bar
h^{(0)}$ are (+) branch solutions and the trajectories in region $\bar
h>\bar h^{(0)}$ are (--) branch solutions. Therefore the branch change
always occurs in direction (--)$\rightarrow $ (+) (on upper half $\bar
h=\bar h^{(0)}$ for trajectories with $X<0$ and on lower half $\bar h=-\bar
h^{(0)}$ for trajectories with $X>0$).

We now turn to the analysis of (+)/(--) branch change on phase plane $(\phi
,X)$. In terms of these variables the (+)/(--) branch change curve is
described by the equation (\ref{xtrans1}). In Fig.1 this curve is shown by
dash-dot line for the case $f_1>0$. From $X^{(0)}\geq \sqrt{-\bar V}$ it
follows that it always lies in classically allowed region. Though the phase
trajectories intersect the curve (\ref{xtrans1}) twice (except the
trajectories touching at points $A$ and $B$ ) the branch change occurs only
at one of these points. Namely, the branch change occurs when the
trajectories with $\bar h<0$ (dashed lines in Fig.1, recall that we consider
the case $f^{\prime }<0$) intersect the upper half $X=X^{(0)}$ of curve (\ref
{xtrans1}), and then the trajectories with $\bar h>0$ (solid lines)
intersect the lower half $X=-X^{(0)}$. It can be easily seen that the
trajectories in region $-X^{(0)}<X<-\sqrt{-\bar V}$ ($\sqrt{-\bar V}%
<X<X^{(0)}$) correspond to (+) ((--)) branch solutions. In the region $%
\left| X\right| >X^{(0)}$ the trajectories with $\bar h>0$ (solid lines)
correspond to (--) branch and the trajectories with $\bar h<0$ (dashed
lines) correspond to (+) branch solutions. We see again that branch change
occurs in direction (--)$\rightarrow $ (+).

\section{Conclusions}

We have considered the graceful exit problem from a superinflationary
pre-big bang phase to a decelerated post-big bang one within the context of
lowest order string effective action, when higher genus terms of general
form and additional matter fields are present. The chosing of the E-frame
essentially simplifies the consideration. We have shown that above mentioned
phase transition does not occur when the conditions (\ref{cond1}), (\ref
{cond2}) are satisfied. The branch change, if it takes place, must be always
in opposite, (--) $\rightarrow $(+) direction. As it follows from here, the
possibility of successful branch change from pre-big bang phase to post
big-bang one within the framework of string effective action (\ref{effact})
requires the violating at least of one of these conditions. Indeed, in \cite
{cal1} it is shown that in the case of Damour-Polyakov ansatz (\ref{prev3})
for the functions $F_k$, continuous (+) $\rightarrow $ (--) branch changing
solutions exist in the region there $B(\phi )<0$. However, as it was
mentioned above, in E-frame corresponding solutions are singular. In Sec.4
the phase space analysis of various approaches to the problem is carried
out. The results are given in Fig.1,2.

\newpage\

\newpage
\centerline{\bf Figure Captions} \vskip .1in {\bf Fig. 1.} The phase
portrait of dynamical system (\ref{dynsys}) on phase space $(\phi ,X)$ for
the potential (\ref{quadpot}) in the case (a) of (\ref{cases}). Solid/dashed
lines correspond to the E-frame expansion/contraction models. $A_1B_1$is one
of the special trajectories with asymptotic behavior (\ref{sptraj}). The
dash-dot lined curve correspond to the string frame (+)/(--) branch change
curve (\ref{xtrans1}). The (--) $\rightarrow $ (+) branch change occur when
the dashed lines intersect the upper half of this curve or when solid lines
intersect the lower half.

{\bf Fig.2.} The phase portrait of E-frame dynamical system (\ref{dynsyse})
with potential (\ref{quadpot}) on phase space $(\phi ,h)$ . The classically
forbidden region is shaded. The trajectories passing through points $B_1$
and $B_2$ present special trajectories (\ref{sptraj}). The dash - dot line
corresponds to the string frame branch changing curve (\ref{hubeint}). (--)$%
\rightarrow $(+) branch change occurs when the solid line trjectories
intersect lower half of this curve or when the dashed lines intersect the
upper half.


\begin{thebibliography}{99}
\bibitem{infl}  A.D.Linde,{\it Particle Physics and Inflationary Cosmology},
(Harwood Academic Publishers, New York, 1990);\\K.A.Olive, Phys.Rep., {\bf %
190}, 307 (1990).

\bibitem{oldinfl}  A.H.Guth, {\it Phys.Rev.} {\bf D23}, 347 (1981).

\bibitem{Guth}  A.H.Guth and E.J.Weinberg, {\it Nucl. Phys.} {\bf B212}, 321
(1983).

\bibitem{newinfl}  A.D.Linde, {\it Phys. Lett.} {\bf B108}, 389 (1982);\\%
A.Albrecht and Steinhardt, {\it Phys. Rev. Lett.} {\bf 48}, 1220 (1982).

\bibitem{chaotic}  A.D.Linde, {\it Phys. Lett.} {\bf B219}, 177 (1983).

\bibitem{extinfl}  D.La and P.J.Steinhardt, {\it Phys.Rev.Lett.} {\bf 62},
376 (1989); {\it Phys.Lett.} {\bf B220}, 375 (1989).

\bibitem{powerlaw}  C.Mathiazhagen and V.B.Johri, {\it Class. Quantum Grav.} 
{\bf 1}, L29 (1984)\\B.~L.~Spokoinyi, {\it Phys. Lett.} {\bf B147}, 39 (1984)%
\\M.~D.~Pollock, {\it Phys. Lett.} {\bf B156}, 301 (1985); {\it Nucl. Phys.} 
{\bf B277}, 513 (1986)\\F.~Accetta, D.~J.~Zoller and M.~S.~Turner, {\it %
Phys. Rev.} {\bf D31}, 3046 (1985)\\F.~Lucchin, S.~Matarrese and
M.~D.~Pollock, {\it Phys. Lett.} {\bf B167}, 163 (1986).

\bibitem{exppot}  L.~F.~Abbott and M.~B.~Wise, {\it Nucl. Phys.} {\bf B244},
541 (1984)\\F.~Lucchin and S.~Matarrese, {\it Phys. Rev.} {\bf D32}, 1316
(1985).

\bibitem{extinfl1}  E.~J.~Weinberg, {\it Phys. Rev.} {\bf D40}, 3950 (1989)\\%
D.~La, P.~J. ~Steinhardt and E.~Bertschinger, {\it Phys. Lett.} {\bf B231},
231 (1989).

\bibitem{hypext}  F.~S.~Accetta and J.~J.~Trester, {\it Phys. Rev.}{\bf D39}%
, 3854 (1989)\\P.~J.~Steinhardt and F.~S.~Accetta, {\it Phys. Rev. Lett.} 
{\bf 64}, 2740 (1990)\\R.~Holman, E.~W.~Kolb and Y.~Wang, {\it Phys. Rev.
Lett.} {\bf 65}, 17 (1990)\\J.~D.~Barrow and K.-I.~Maeda, {\it Nucl. Phys.} 
{\bf B341}, 294 (1990).

\bibitem{multinf}  R.Holman, E.~W.~Kolb, S.~L.~Vadas and Y.~Wang, {\it Phys.
Rev.} {\bf D43}, 995 (1990)\\A.~S.~Majumdar and S.~K.~Sethi, {\it Phys. Rev.}
{\bf D46}, 5315 (1992).

\bibitem{kalara}  S.~Kalara, N.~Kaloper and K.~A.~Olive, {\it Nucl. Phys.} 
{\bf B341}, 252 (1990).

\bibitem{hybridinfl}  A.~D.~Linde, {\it Phys. Lett.} {\bf B259}, 38 (1991)\\%
A.~R.~Liddle and D.~H.~Lith, {\it Phys. Rep.} {bf 231}, 1 (1993)\\%
A.~D.~Linde, {\it Phys. Rev.} {\bf D49}, 748 (1994)\\E.~J.~Copeland,
A.~R.~Liddle, D.~H.~Lith, E.~D.~Stewart and D.~Wands, {\it Phys. Rev.} {\bf %
D49}, 6410 (1994)\\A.~D.~Linde and A.~Mezhlumian, {\it Phys. Rev.} {\bf D52}%
, 6789 (1995)\\J.~Garcia-Bellido and A.~D.~Linde, ''Tilted hybrid
inflation'', CERN-TH/96-358, astro-ph/9612141; ''Open hybrid inflation'',
CERN-TH/97-08, astro-ph/9701173.

\bibitem{softinf}  A.~L.~Berkin, K.~Maeda and J.~Yokoyama, {\it Phys. Rev.
Lett.} {\bf 65}, 141 (1990)\\A.~L.~Berkin and K.~Maeda, {\it Phys. Rev.} 
{\bf D44}, 1691 (1991).

\bibitem{natinf}  L.~Randall, M.~Soljacic and A.~H.~Guth, {\it Nucl. Phys.} 
{\bf B472}, 349 (1996).

\bibitem{supinf}  P.~Binetruy and M.~K.~Gaillard, {\it Phys. Rev.} {\bf D34}%
, 3069 (1986)\\B.~A.~Campbell,A.~D.~Linde and K.~A.~Olive, {\it Nucl. Phys.} 
{\bf B355}, 146 (1991)\\M.~C.~Bento, O.~Bertolami and P.~M.~Sa, {\it Phys.
Lett.} {\bf B262}, 11 (1991)\\S.~Thomas, preprint SLAC-PUB-95-6767.\\%
T.~Banks, M.~Berkooz,S. H. Shenker, G. Moore and P. J. Steinhardt, {\it %
Phys. Rev.} {\bf D52}, 3452 (1995);\\J. Garcia-Bellido and M. Quiros, {\it %
Nucl. Phys.} {\bf B368}, 463 (1992);\\M. C. Bento and O. Bertolami, preprint
CERN-TH/95-199;\\G. G. Ross and S. Sarkar, {\it Nucl. Phys.} {\bf B461}, 597
(1996);\\P. Binetruy and G. Dvali, preprint CERN-TH/96-149;\\G. Dvali,
preprint CERN-TH/96-129;\\T. Damour and A. Vilenkin, {\it Phys. Rev.} {\bf %
D53}, 2981 (1996),\\A.A.Saharian, {\it Astrophysics }{\bf 38}, 101 (1995); 
{\bf 39, }153 (1996).

\bibitem{sah1}  A. A. Saharian, {\it Astrophysics}, {\bf 38}, 447 (1995).

\bibitem{sdual}  G. Veneziano, {\it Phys. Lett.} {\bf B265}, 287 (1991);\\A.
Tseytlin, {\it Mod. Phys. Lett.} A, 1721 (1991).

\bibitem{gasp1}  M. Gasperini and G. Veneziano, {\it Astropart. Phys.} {\bf 1%
}, 317 (1993).

\bibitem{gasp2}  M. Gasperini and G. Veneziano, {\it Mod. Phys. Lett.} {\bf %
A8}, 3701 (1993).

\bibitem{gasp3}  M. Gasperini and G. Veneziano, {\it Phys. Rev.} {\bf D50},
2519 (1994).

\bibitem{levin}  J. Levin, {\it Phys. Rev.} {\bf D51}, 1536 (1995).

\bibitem{brust}  R. Brustein and G. Veneziano, {\it Phys. Lett.} {\bf B329},
429 (1994).

\bibitem{cal1}  N. Caloper, R. Madden and K. A. Olive, {\it Nucl. Phys.} 
{\bf B452}, 677 (1995).

\bibitem{cal2}  N. Caloper, R. Madden and K. A. Olive, {\it Phys. Lett.} 
{\bf B371} 34 (1996).

\bibitem{east}  R. Easter, K. Maeda and D. Wands, {\it Phys. Rev.} {\bf D53}%
, 4247 (1996).

\bibitem{quantcos}  M. Gasperini, J. Maharana and G. Veneziano, {\it Nucl.
Phys.} {\bf B472}, 349 (1996);\\J. E. Lidsey, ''Inflationary and
deflationary branches inextended pre-big bang cosmology'', gr-qc/9605017.

\bibitem{higher}  R. Easter and K. Maeda, ''One-loop superstring cosmology
and the non-singular universe'', hep-th/9605173;\\S. J. Rey, {\it Phys. Rev.
Lett.} {\bf 77}, 1929 (1996);\\M. Gasperini and G. Veneziano, {\it Phys.
Lett.} {\bf B387}, 715 (1996).

\bibitem{staction}  C. Lovelace, {\it Phys. Lett.} {\bf B135}, 75 (1984);\\%
C. G. Callan, D. Friedan, E. J. Martinec and M. J. Perry, {\it Nucl. Phys.} 
{\bf B262}, 593 (1985);\\E. S. Fradkin and A. A. Tseytlin, {\it Phys. Lett.} 
{\bf B158}, 316 (1985); {\it Nucl. Phys.} {\bf B261}, 1 (1985);\\A. Sen, 
{\it Phys. Rev.} {\bf D32}, 2102 (1985); {\it Phys. Rev. Lett.} {\bf 55},
1846 (1985);\\C. G. Callan, I. R. Klebanov and M. J. Perry, {\it Nucl. Phys.}
{\bf B278}, 78 (1986);\\D. J. Gross and J. H. Sloan, {\it Nucl. Phys.} {\bf %
B291}, 41 (1987);\\J. Lauer, D. Luest and S. Theisen, {\it Nucl. Phys.} {\bf %
B304}, 236 (1988).

\bibitem{copel}  E. J. Copeland, A. Lahiri and D. Wands, {\it Phys. Rev.} 
{\bf D50}, 4868 (1994).

\bibitem{sahpot}  A. A. Saharian, {\it Astrophysics} {\bf 40, }153 (1997); 
{\bf 40,} No.3 (1997) (in press).

\bibitem{damour}  T. Damour and A. M. Polyakov, {\it Nucl. Phys.} {\bf B423}%
, 532 (1994).

\bibitem{hawking}  S. W. Hawking and G. F. R. Ellis, {\it The large scale
structure of space-time}. (Cambridge University Press, 1973).
\end{thebibliography}
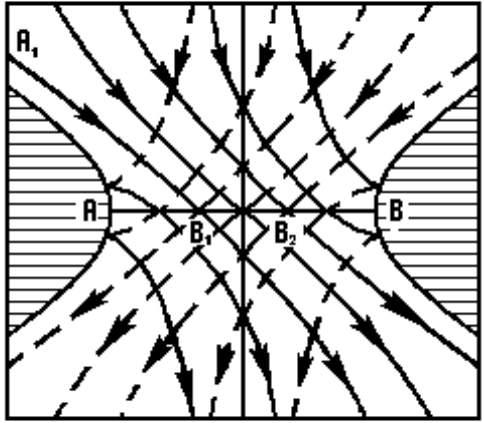
\end{document}